\documentclass[prb,twocolumn]{revtex4}

\usepackage{graphicx}
\usepackage{verbatim}
\begin{document}

\def\K{{\bf{K}}}
\def\Q{{\bf{Q}}}
\def\Gbar{\bar{G}}
\def\tk{\tilde{\bf{k}}}
\def\k{{\bf{k}}}

\title{Effect of long-range hoppings on $T_c$ in a two-dimensional
   Hubbard-Holstein model of the cuprates}
\author{E. Khatami}
\author{A. Macridin}
\author{M. Jarrell} \affiliation{Department of Physics, University of 
Cincinnati, Cincinnati, Ohio, 45221, USA} 


\begin{abstract}

We study the effect of long-range hoppings on $T_c$ for the 2D Hubbard 
model with and without Holstein phonons using parameters evaluated from 
band structure calculations for cuprates. Employing the dynamical cluster 
approximation (DCA) with a quantum Monte Carlo (QMC) cluster solver for a
4-site cluster, we observe that without phonons, the long-range hoppings 
$t'$ and $t''$, generally suppress $T_c$. We argue that this trend remains
valid for larger clusters.  In the presence of the Holstein phonons, a 
finite $t'$ enhances $T_c$ in the under-doped region for the hole-doped 
system, consistent with LDA calculations and experiment. This is interpreted 
through the suppression of antiferromagnetic correlations and the interplay
between polaronic effects and the antiferromagnetism.

\end{abstract}

\date{\today}

\maketitle


\paragraph*{Introduction-}

While most theoretical studies of the cuprates are in the framework of 
the simplest version of the two-dimensional Hubbard model with only 
nearest neighbor hopping, both band structure calculations and experimental 
data suggest a richer set of parameters for this model
~\cite{params,anderson,k_tanaka_04,a_nazarenko_95,c_kim_98,v_belinicher_96,r_eder_97,f_lema_97}. 
Angle-resolved photoemission spectroscopy (ARPES) plays an important role in 
this regard, suggesting different topologies of the Fermi surface for 
different high-$T_c$ superconductors~\cite{k_tanaka_04,a_damascelli_03} 
which can be reproduced by choosing finite long-range hoppings ($t', t'', 
\dots$) \cite{a_nazarenko_95,c_kim_98,v_belinicher_96,r_eder_97,f_lema_97}. 
The inclusion of the next nearest neighbor hopping, $t'$, in the Hubbard 
model is also necessary to capture the electron-hole asymmetry 
~\cite{t_tohyama_04,g_martins_01,r_gooding_94}.  Furthermore, $t'$ is an 
important parameter in determining the charge orderings and their 
textures in cuprates~\cite{s_white_98,g_seibold_06,g_zha_07,a_himeda_02}.

The effect of $t'$ on $T_c$ has been studied by different groups
~\cite{r_raimondi_96,anderson,k_tanaka_04}. For example, E. Pavarini 
{\em et.al.}\cite{anderson} notice a correlation between the experimental 
maximum superconducting temperature ($T_c^{max}$) and the value 
of $t'$ evaluated from the band structure calculations in different cuprates. 
However, the mechanism which may govern this relationship in cuprates is 
not well-understood. 

Theoretical investigations employing simple models such as single-band 
Hubbard and t-J models~\cite{shih,l_Spanu_08,x_chen_04,x_chen_05,g_martins_01}, do 
not show strong evidence of a direct relationship between $T_c$ in 
different doping regions and the magnitude of the long-range hoppings. 
Variety of techniques have been used to study the effect of $t'$ and 
$t''$ (third nearest neighbor hopping) on superconducting properties of 
these models.  For hole-doped systems, by employing finite size calculations 
and slave-boson mean field theory, C. T. Shih {\em et. al.}~\cite{shih} 
find a strong enhancement of the superconducting correlations due to $t'$ 
and $t''$ in the intermediate- and over-doped regions and a slight suppression 
in the under-doped region.  For electron-doped systems, density matrix 
renormalization group calculations~\cite{s_white_98} have shown that
$t'$ leads to the enhancement of d-wave pairing correlations. 
Unlike  finite-size calculations where the transition temperature cannot 
be directly calculated and the superconducting properties are estimated 
from cluster  pairing correlations, the DCA~\cite{m_hettler_98,m_hettler_00,algorithm} 
is an approximation for the thermodynamic limit and allows calculation of 
$T_c$ and other superconducting properties for all doping regions.
The DCA has been successful to derive the phase diagram of the Hubbard 
model, showing the antiferromagnetic (AF), pseudogap and d-wave 
superconductivity phases which are in a very good qualitative agreement
with experiments~\cite{th_maier_05,Alex}.

In this work, we investigate the effect of $t'$ and $t''$ on the 
superconducting properties of the 2D single-band Hubbard model with and 
without phonons.  We find that without phonons, $T_c$ is generally 
suppressed by $t'$ and $t''$. However, with Holstein phonons, $T_c$ 
increases with $t'$ in the under-doped region for hole-doped systems. 
In other doping regions, phonons reduce the suppression of $T_c$ due to $t'$.
The interplay between electron-phonon (EP) coupling and $t'$ plays an 
essential role in the dependence of $T_c$ on $t'$ and is consistent with 
experimental data showing evidence of strong EP interaction in the 
cuprates~\cite{a_lanzara_01,t_cuk_05}.  Previously, we 
established a synergistic relationship between short ranged AF order and 
EP coupling in the doped Hubbard model~\cite{a_macridin_06}, {\em i.e.} we 
found that  AF correlations enhance the polaronic effects 
(dressing of electrons by phonons) and at the 
same time, the EP coupling enhances the AF correlations.  We also established
that local phonons which couple to the electronic density,
strongly suppress $T_c$ due to the renormalization of the single-particle 
propagator.  Here, we show that $t'$ can strongly affect this synergism 
and thus the suppression of $T_c$. A finite $t'$ in the hole-doped systems, 
 suppresses AF correlations and hence 
reduces the polaronic effects and enhances $T_c$.


\paragraph*{Model-}
\label{sec:Model}

We consider a 2D Hubbard-Holstein model
\begin{eqnarray}
\label{eq:1}
H&=&-\sum_{ij\sigma}t_{ij}(c^{\dagger}_{{i}\sigma}c_{{j}\sigma}+h.c.)+
\epsilon\sum_{{i}\sigma}n_{{i}\sigma}+U\sum_{i}n_{{i}
\uparrow}n_{{i}\downarrow} \nonumber \\
&+&\sum_i \frac{p_i^2}{2M}+\frac{1}{2}M\omega_0^2 u_i^2+gn_i u_i
\end{eqnarray}
where $t_{ij}$ is the hopping matrix, $c^{\dagger}_{{i}\sigma}(c_{{i}\sigma})$ 
is the creation (annihilation) operator for electrons on site ${i}$ with spin 
$\sigma$, and $U$ is the on-site Coulomb repulsion which is taken to be equal to 
the bandwidth ($8t$). We vary the filling, $\langle n \rangle$, from values less 
than one to values larger than one to cover the hole-doped to the 
electron-doped regions respectively. $\omega_0$ is the frequency of phonons 
and $\{u_i, p_i\}$ are canonical conjugate coordinates for the phonon on 
site $i$. The EP coupling is on-site and proportional to the density of 
electrons with the coupling strength $g$. We define the dimensionless 
EP coupling for Holstein phonons as 
\begin{equation}
\label{eq:lambda}
\lambda=g^2/(M\omega_0^28t) 
\end{equation}
which is the ratio of the single-electron lattice deformation energy and half 
of the electronic bandwidth\cite{c_slezak_06}.


\paragraph*{Formalism-}
\label{sec:formalism}

We employ the dynamical cluster approximation with a quantum Monte Carlo 
algorithm as the cluster solver. 
The DCA approximates the self-energy 
of the system by mapping it into a cluster of size $N_c$ 
embedded in a self consistent host. All of the correlations inside 
the cluster are treated non-perturbatively while a mean-field (MF) 
approximation is used to deal with longer range correlations. 
Therefore, the solution would be exact in the limit of $N_c\rightarrow\infty$.
The Monte Carlo simulation  performs the sum over both the
discrete field used to decouple the Hubbard repulsion\cite{jarrell:dca}, as well as the
phonon field, $u$.  Details about an efficient  Monte Carlo simulation of
 systems with low energy
phonons  will be given elsewhere.

The sign problem in QMC limits our calculations to relatively small clusters.
Most of the calculations are done for a $2\times2$ cluster, 
the smallest cluster which allows 
$d-$wave pairing. Note that the length scale associated with $t''$ is not
represented for this cluster. Thus, the effect of $t''$ 
will be similar to that in the dynamical mean field approximation~\cite{georges,Pruschke};
i.e. only through changes in the cluster densities of states.

\begin{figure}[t]
\leftline {\includegraphics*[width=3.24in]{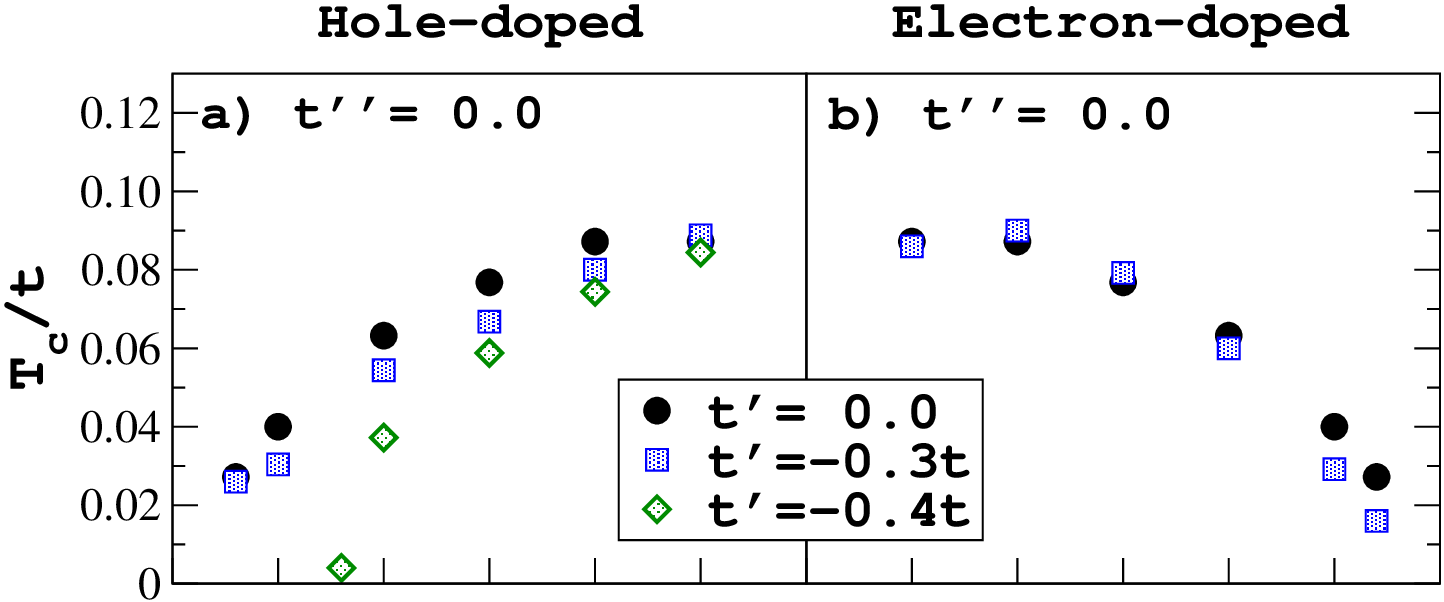}}
\leftline {\includegraphics*[width=3.3in]{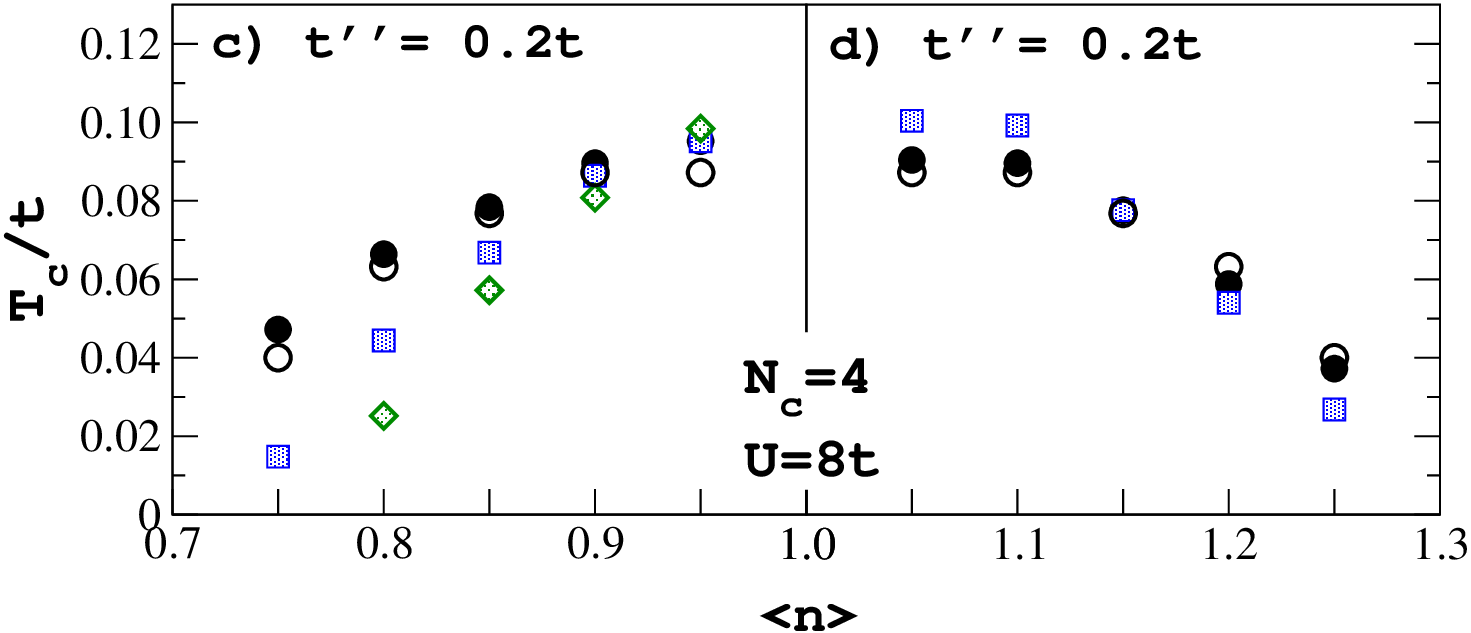}}
\caption{(color online) The superconducting phase diagram of the 
hole-doped (left panels) and the electron-doped (right panels) system
for different values of $t'$. Upper (lower) panels correspond to 
$t''=0$ ($t''=0.2t$). The open circles in c) and d) correspond 
to $t'=t''=0$, and are plotted for comparison.}
\label{fig:Tcn}
\end{figure}

\begin{figure}[t]
\centerline {\includegraphics*[width=3.3in]{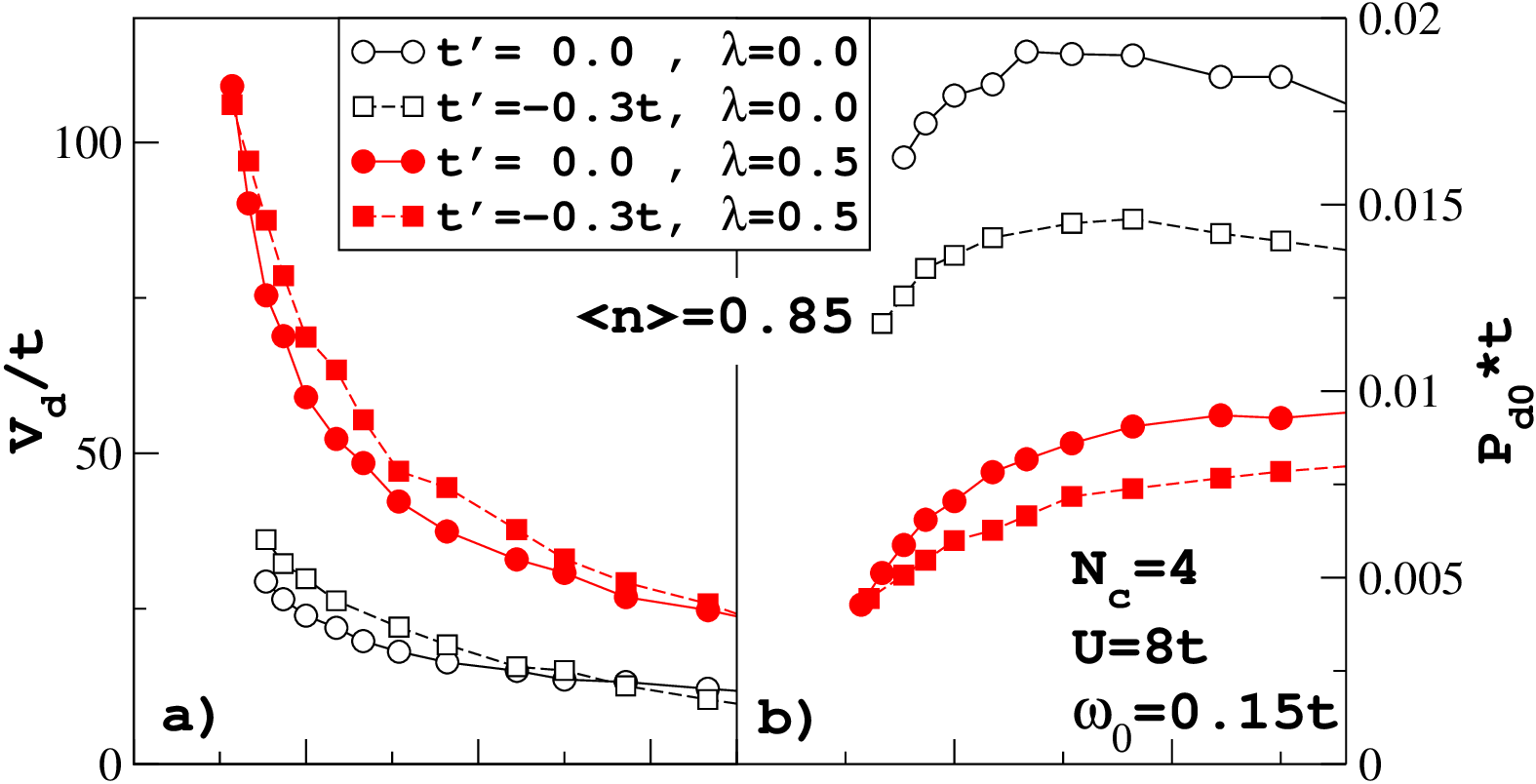}}
\centerline {\includegraphics*[width=3.3in]{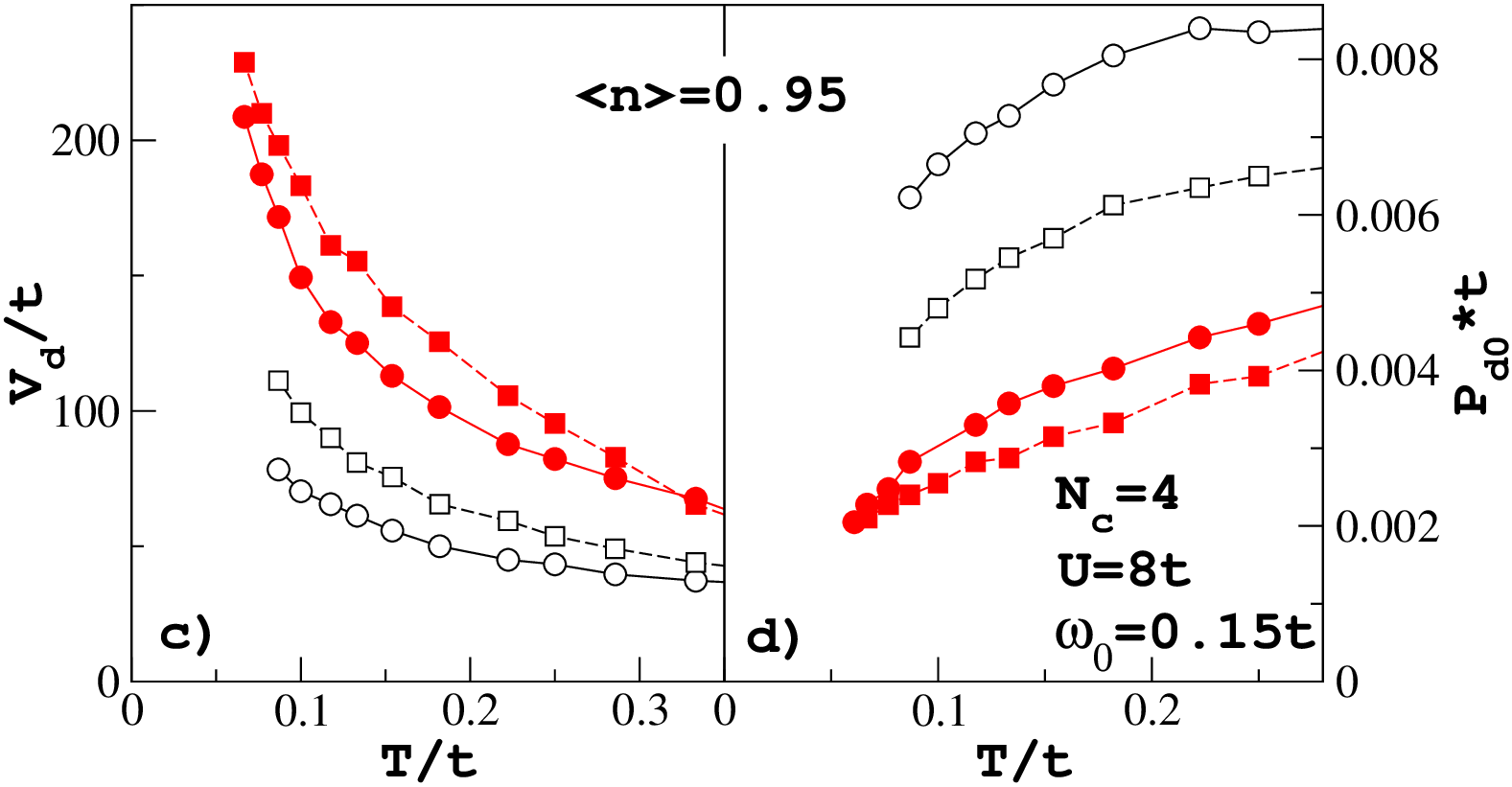}}
\caption{(color online) $d-$wave pairing interaction, $V_d$ (Eq.~\ref{eqn:Vd}), and the d-wave 
projected bare bubble, $P_{d0}$ (Eq.~\ref{eqn:Pd0}), versus temperature
in the under-doped and intermediate-doped regions. 
The empty (filled) symbols correspond to the Hubbard model without 
(with) Holstein phonons. Circles (squares) correspond to $t'=0.0$ ($t'=-0.3t$).}
\label{fig:Vd}
\end{figure}

In order to investigate the effect of $t'$ on the d-wave 
superconductivity,  we calculate the eigenvalues of the pairing matrix
$\Gamma{\chi_0}$, where ${\chi_0}$ is the bare bubble and $\Gamma$ is 
the particle-particle irreducible vertex function calculated in the QMC process. 
At $T_c$, the leading eigenvalue (in this case, the one with d-wave symmetry) goes 
to unity and causes a singularity in the two-particle pairing Green's function
${\chi}=\chi_0 +\chi_0 \Gamma \chi ={\chi_0}/(1-\Gamma{\chi_0})$. 
The value of the d-wave pairing interaction can be measured by calculating
the d-wave projected vertex~\cite{th_maier_06,th_maier_07}
\begin{equation}
\label{eqn:Vd}
V_d=-\frac{\langle g(\K)\Gamma(\K,\pi T|\K',\pi T)g(\K')\rangle_{\K\K'}}
{\langle g(\K)^2\rangle _{\K}}
\end{equation}
for the lowest Matsubara frequency and $g(\K)=cos(\K_x)-cos(\K_y)$
where $\K$ is the momentum at the center of each of the $N_c$ cells 
which tile the Brillouin Zone in the DCA.
To capture the effect of the dressed electronic propagator on the 
d-wave eigenvalue, we also calculate the d-wave projected bare bubble as
\begin{equation}
\label{eqn:Pd0}
P_{d0}=\frac{T}{N_c}\frac{\langle g(\K)^2\chi_0(\K,\pi T)\rangle_{\K}}
{\langle g(\K)^2\rangle _{\K}}
\end{equation}


\paragraph*{Results-}
\label{sec:results}

The long-range hoppings, $t'$ and $t''$ can affect the superconducting
phase diagram through both the band structure and the interaction vertex.
In the electron-doped systems, $t'$ favors hopping in the same sub-lattice and
enhances the AF correlations at finite doping while in the hole-doped systems, 
$t'$ suppresses the AF correlations~\cite{t_tohyama_94,t_tohyama_04}. 
Presumably, $t''$ which also introduces hopping
in the same sub-lattice, would affect AF correlations as well. However, 
previous calculations indicate that there is a close relationship between 
AF and superconductivity in cuprates~\cite{th_maier_06b,th_maier_07,th_maier_07b}. 
Therefore, $t'$ and $t''$ can influence pairing by affecting the AF 
correlations. They also change the band structure which alters the density 
of states at the Fermi energy and thus can influence $T_c$.

We find that $t'$ and $t''$ generally suppress $T_c$ in most doping
regions, apart from a slight increase in $T_c$ at small dopings.
First, we consider a finite $t'$ and $t''=0$. The superconducting 
phase diagrams for three different values of $t'$ are shown in 
Fig.~\ref{fig:Tcn} (a) and (b). When $t'=-0.3t$, $T_c$ is slightly smaller 
in comparison with the case of $t'=0$ from 10\% to 25\% hole-doping 
and at large electron-doping ($\geq$20\%). With a larger $t'=-0.4t$ for  
the hole-doped system, we find that $T_c$ is strongly suppressed in 
intermediate- and over-doped regions while in the under-doped region, 
the overall effect of $t'$ on $T_c$ is negligible. The effect of 
$t''$ on the superconducting phase diagram is shown in Fig.~\ref{fig:Tcn} 
(c) and (d). We see a stronger suppression of $T_c$ in the over-doped region 
and a slight increase ($\sim 10\%$) in $T_c$ in the under-doped region 
when both $t'$ and $t''$ are finite. Moreover, a non-zero $t''$ with 
$t'=0$, does not have a considerable effect on $T_c$~\cite{footnote}.

\begin{figure}[t]
\centerline {\includegraphics*[width=3.3in]{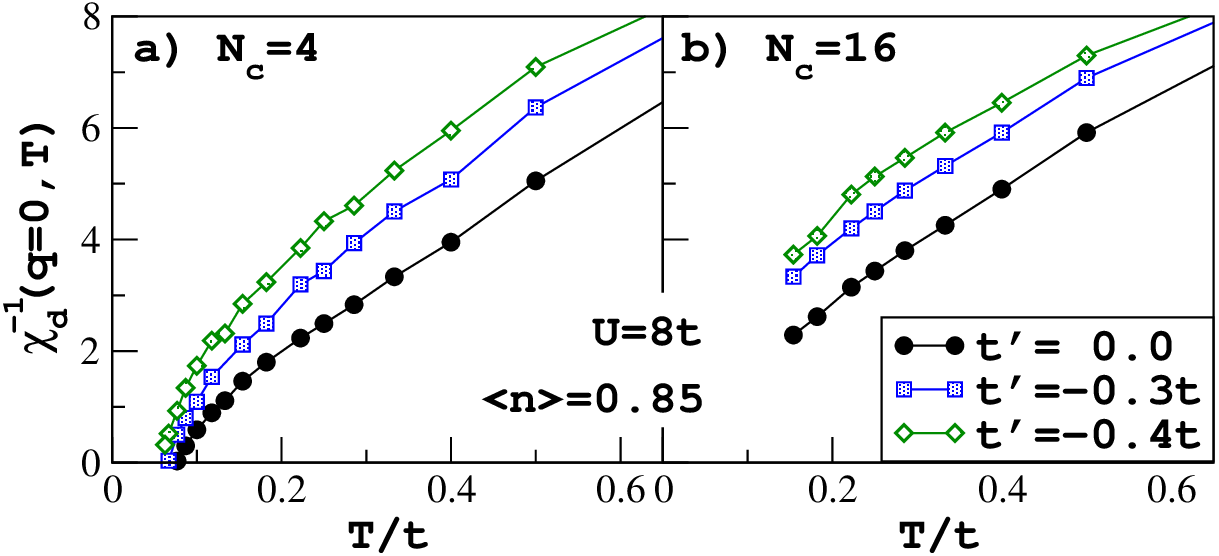}} 
\caption{(color online) Inverse of the d-wave pairing susceptibility for a) 
4-site and b) 16-site cluster at 15\% doping for different next-nearest hoppings.}
\label{fig:16-site}
\end{figure}

We find that the band renormalization effects due to $t'$ are mostly 
responsible for changes in $T_c$ and the effect of $t'$ on the interaction 
vertex is less significant. 
In order to illustrate this, we plot the $t'$-dependence 
of the d-wave bare bubble, $P_{d0}$ (Eq.~\ref{eqn:Pd0}), and the 
d-wave pairing interaction, $V_d$ (Eq.~\ref{eqn:Vd}), at
5\% and 15\% hole-doping in Fig.~\ref{fig:Vd} (empty symbols). $t'$ 
strongly suppresses $P_{d0}$ and slightly increases $V_d$ in both 
doping regions. The former effect is responsible for the decrease 
in $T_c$. The suppression in $P_{d0}$ is a result of the band 
renormalization effects of $t'$ which decrease the density of 
states at the anti-nodal points~\cite{t_tohyama_94,a_macridin_06b}. 

The effect of $t'$ on the d-wave pairing shows a similar trend when 
larger clusters are considered. The inverse of the d-wave pairing
susceptibility for 4-site and 16-site clusters at 15\% 
hole-doping are shown in Fig.~\ref{fig:16-site}. For both clusters 
and in the temperature range available, $t'$ suppresses 
the d-wave pairing susceptibility. 

In the presence of phonons, long-range hoppings change the 
polaronic effects which have a direct influence on $T_c$.
In previous works, we have found that at the intermediate EP coupling,
local phonons suppress $T_c$ due 
to  polaronic effects which reduce the mobility of carriers~\cite{Alex}.
The polaronic effects are enhanced by the AF correlations~\cite{j_zhong_92}.
Therefore, the effect of $t'$ on the AF correlations will directly influence 
them.  Using exact diagonalization methods, T. Tohyama 
{\em et. al.}~\cite{t_tohyama_94,t_tohyama_04} have shown that $t'$ 
suppresses the AF correlations in the hole-doped cuprates.  As a result, 
the polaronic effects are reduced by $t'$ which enhances $T_c$.

\begin{figure}[t]
\centerline {\includegraphics*[width=3.3in]{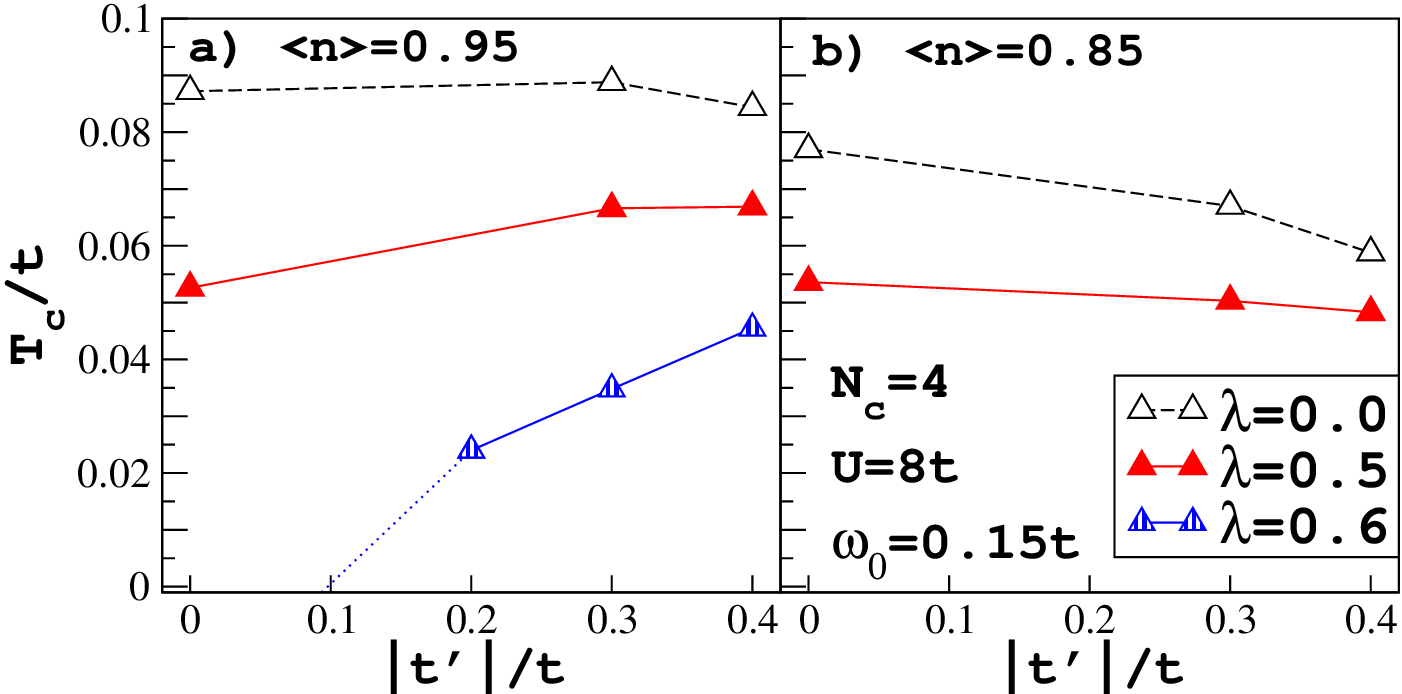}} 
\caption{(color online) $t'$-dependence of $T_c$ 
at a) 5\% and b) 15\% dopings for different values of the dimensionless EP coupling, 
$\lambda$ (Eq.~\ref{eq:lambda}). A finite $t'$ increases $T_c$ in the 
under-doped region when EP coupling is present.
For $\lambda=0.6$ in (a), the $t'=0$ point has been extrapolated form
the data at temperatures larger than $T_c$.}
\label{fig:Tc}
\end{figure}

When Holstein phonons are considered,  $T_c$ increases with $t'$ in 
the under-doped region and remains almost unchanged in the intermediate-doped 
region as shown in Fig.~\ref{fig:Tc}.  Note that without phonons, $T_c$ 
decreases  with $t'$ in the intermediate-doped and does not change in the 
under-doped region. The effect of $t'$ on $T_c$ becomes more significant
with increasing $\lambda$. As shown in Fig.~\ref{fig:Tc} (a),
for $\lambda=0.6$ at 5\% doping, $T_c$ is strongly enhanced when 
$|t'|$ increases. At 15\% doping, it is 
difficult to fix the filling due to the large charge fluctuations 
for the values of $\lambda$ larger than $0.5$. 
Presumably, $T_c$ would increase with $t'$ at 15\% doping 
for $\lambda=0.6$. Note that phonons do not increase $T_c$, they only reverse 
the behavior of $T_c$ with $t'$.  At fixed $t'$, their effect is to 
reduce  $T_c$, but the reduction is less significant
when $|t'|$ is larger.

Phonons change the behavior of $P_{d0}$ with respect to $t'$
at low temperatures. As shown in  
Fig.~\ref{fig:Vd} (b) and (d) with full symbols,
$t'$ has a small effect on suppressing $P_{d0}$
around $T_c$ when phonons are present. While the band 
renormalization effects, caused by $t'$, 
tend to suppress $P_{d0}$, the reduction in the polaronic effects  
due to $t'$ enhances $P_{d0}$. As a result of these two 
competing effects, $P_{d0}$ remains almost unchanged 
near $T_c$ by changing $t'$. 
On the other hand, the $t'$-dependence of $V_d$
is not influenced much by phonons. Therefore, 
$T_c$ increases in the under-doped region where $V_d$
shows a slight increase with $t'$.

It is known that the $2 \times 2$ cluster  overestimates the  d-wave
superconductivity due to the neglect of 
phase fluctuations~\cite{th_maier_05b}. However, here we focus mainly on investigating
the relative dependence  of $T_c$ on different parameters such as EP coupling and long-range hoppings
and not on calculating the exact value of $T_c$.  The $16$-site cluster results
(shown in Fig.~\ref{fig:16-site})  suggest that these trends do not  change
when larger clusters are considered. 

The transition temperature at small doping would be 
the strongest affected by phase fluctuations~\cite{phase}.  
In general, $t'$  should enhance $T_c$ by suppressing the phase fluctuations.  In part, this 
is due to the suppression of AF correlations. The AF correlations reduce 
the mobility of the carriers and increase the effective mass which leads to the 
enhancement of phase fluctuations~\cite{phase}. This effect 
is enhanced in the presence of phonons. Phonons play a role similar to the 
AF correlations in enhancing the phase fluctuations by reducing the 
mobility of electrons.  In addition, through polaronic effects, they also 
enhance the AF correlations. Therefore, the effect of phonons on suppressing 
$T_c$ is underestimated in the absence of the phase fluctuations. 
The decrease of the polaronic effects due to $t'$ would increase $T_c$ 
more significantly in the under-doped region in the presence of phase 
fluctuations. Hence, the effect of $t'$ on $T_c$ is underestimated by small 
cluster calculations, suggesting that $t'$ will have a stronger effect on 
$T_c$ in larger clusters.


\paragraph*{Conclusion-}

We find that without phonons, the long-range hoppings generally 
suppress $T_c$ in the Hubbard model. However, by including Holstein phonons, $T_c$ 
increases with $t'$ in the under-doped region for the hole-doped system 
while the suppression in $T_c$ due to $t'$ is reduced in the 
intermediate-doped region. Phonons do not increase $T_c$, but rather reverse 
the behavior of $T_c$ with $t'$. 
We find that the increase in $T_c$ with $t'$ becomes 
more significant for larger values of the EP coupling. We interpret these by the 
effect of $t'$ on suppressing the polaronic effects as a result of suppressing 
the AF correlations and the interplay between AF and EP 
coupling. 


\paragraph*{Acknowledgment-}
We acknowledge useful discussions with Th.\ Pruschke.
This research was supported by NSF grant Nos. DMR-0706379 and DMR-0312680,
and DOE CMSN DE-FG02-04ER46129 and enabled by allocation of advanced 
computing resources, supported by the National Science Foundation. 
The computations were performed in part on Lonestar at the Texas 
Advanced Computing Center (TACC) under account No. TG-DMR070031N.
Part of this research was enabled by resources in Ohio Supercomputer 
Center under the project No. PES0609.



\begin{thebibliography}{99}
\bibitem{footnote} The effect of $t'=-0.4t$ for the electron-doped system was not 
explored due to the computational expense. For this reason, from now on, 
we focus on the effect of $t'$ only in the hole-doped system.
\end{thebibliography}
\end{document}